\documentclass{article}
\usepackage{spconf,amsmath,epsfig,amssymb,amsthm}

\def\x{{\mathbf x}}
\def\s{{\mathbf s}}
\def\y{{\mathbf y}}
\def\H{{\mathbf H}}
\def\HH{\hat{\mathbf H}}
\def\V{{\mathbf V}}

\def\W{{\mathbf W}}
\def\C{{\mathcal C}}

\def\Complex{\mathbb{C}}

\def\Grassmann{\mathbb{G_{\text{M,N}}(\Complex)}}

\def\LSB{\left[}
\def\RSB{\right]}
\def\LB{\left(}
\def\RB{\right)}

\def\ie{i.\@e.\@}

\newtheorem{thm}{Theorem}[section]

\title{MIMO BROADCAST CHANNELS WITH BLOCK DIAGONALIZATION AND FINITE RATE FEEDBACK}
\name{Niranjay Ravindran and Nihar Jindal}
\address{   Department of Electrical and Computer Engineering\\
University of Minnesota\\
Minneapolis, MN 55455, USA\\
Email: \{ravi0022, nihar\}@umn.edu}
\begin{document}
\ninept
\topmargin=0mm

\maketitle
\begin{abstract}
Block diagonalization is a linear precoding technique for the
multiple antenna broadcast (downlink) channel that involves
transmission of multiple data streams to each receiver such that no
multi-user interference is experienced at any of the receivers. This
low-complexity scheme operates only a few dB away from capacity but does
require very accurate channel knowledge at the transmitter, which
can be very difficult to obtain in fading scenarios. We consider a
limited feedback system where each receiver knows its channel
perfectly, but the transmitter is only provided with a finite number
of channel feedback bits from each receiver. Using a random vector
quantization argument, we quantify the throughput loss due to
imperfect channel knowledge as a function of the feedback level. The
quality of channel knowledge must improve proportional to the SNR in
order to prevent interference-limitations, and we show that scaling
the number of feedback bits linearly with the system SNR is
sufficient to maintain a bounded rate loss. Finally, we investigate
a simple scalar quantization scheme that is seen to achieve the same
scaling behavior as vector quantization.
\end{abstract}
\section{Introduction}
\label{sec:intro} In multiple antenna broadcast (downlink) channels,
transmit antenna arrays can be used to simultaneously transmit data
streams to receivers and thereby significantly increase throughput.
Dirty paper coding (DPC) is capacity
achieving for the MIMO broadcast channel \cite{WEIN04},
but this technique has a very high level of complexity.  Zero
Forcing (ZF) and Block Diagonalization (BD) \cite{CHOI04}
\cite{SPEN04} are alternative low-complexity transmission
techniques.  Although not optimal, these linear precoding techniques
utilize all available spatial degrees of freedom and perform
measurably close to DPC in many scenarios \cite{JIND05b}.

If the transmitter is equipped with $M$ antennas and there are at
least $M$ aggregate receive antennas, zero-forcing involves
transmission of $M$ spatial beams such that independent, de-coupled
data channels are created from the transmit antenna array to $M$
receive antennas distributed amongst a number of receivers.  Block
diagonalization similarly involves transmission of $M$ spatial
beams, but the beams are selected such that the signals received at
different receivers, but not necessarily at the different antenna
elements of a particular receiver, are de-coupled.  For example, if
there are $M/2$ receivers with two antennas each, then two beams are
aimed at each of the receivers.  If ZF is used, an independent and
de-coupled data stream is received on each of the $M$ antennas. If
BD is used, the streams for different receivers do not interfere,
but the two streams intended for a single receiver are generally not
aligned with its two antennas and thus post-multiplication by a
rotation matrix (to align the streams) is generally required before
decoding.

In order to correctly aim the transmit beams, both schemes require
perfect Channel State Information at the Transmitter (CSIT).
Imperfect CSIT leads to incorrect beam selection and therefore
multiuser interference, which ultimately leads to a throughput
loss. Unlike point to point MIMO systems where imperfect CSIT causes
only an SNR offset in the capacity vs.\@ SNR curve, the level of CSIT
affects the slope of the curve and hence the \textit{multiplexing}
gain in broadcast MIMO systems. We consider the case when the CSI is
known perfectly at the receiver and is communicated to the
 transmitter through a finite rate feedback channel and  quantify the maximum rate loss due to finite rate
  feedback with BD.
  MISO systems and ZF with finite rate feedback are analyzed in \cite{JIND05}. Similar
   to the results in \cite{JIND05}, we show that scaling the number of feedback bits approximately
   linearly with the system SNR is sufficient to maintain the slope of the capacity vs.\@ SNR curve
    and hence a constant gap from the capacity of BD with perfect CSIT. The scaling factor for
    BD offers an advantage over ZF in terms of the number of bits required to achieve the same
    sum capacity. Finally, we investigate a simple scalar quantization scheme
    and see that this low complexity scheme requires the same
    feedback scaling.

\section{System Model}
\label{sec:sysmodel}

We consider a single transmitter and $K$ user MIMO system where each user has $N$ antennas and the
transmitter has $M$ antennas. The broadcast channel is described as:
\begin{equation}
\y_i = \H_i^H\x + {\bf n}_i,\quad i = 1, \dots, K
\end{equation}
where $\H_i \in \Complex^{M \times N}$ is the channel matrix from the transmitter to the $i^\text{th}$
user ($1 \leq i \leq K$) and the vector $\x \in \Complex^{M \times 1}$ is the transmitted signal.
${\bf n}_i \in \Complex^{N \times 1}$ are independent complex Gaussian noise vectors of unit variance
 and $\y_i \in \Complex^{N \times 1}$ is the received signal vector at the $i^\text{th}$ user. We assume
  a transmit power constraint so that $E[||\x||^2] \leq P$ $(P > 0)$. We also assume that
  $K > 1$ and $K = \frac{M}{N}$, which implies that the aggregate
  number of receive antennas equals the number of transmit antennas;
  as a result it is not necessary to select a subset of receivers
  for transmission.

The entries of $\H_i$ are assumed to be i.\@i.\@d.\@ unit variance complex Gaussian random variables,
and the channel is assumed to be block fading with independent fading from block to block. Each
of the receivers is assumed to have perfect and instantaneous knowledge of their own channel
matrix. The channel matrix is quantized at each receiver and fed back to the transmitter
 (which has no other knowledge of the instantaneous CSI) over a zero delay, error free, finite rate channel.
 In order to perform BD, it is only necessary to know the spatial
 direction of each receiver's channel, i.\@e.\@, the subspace spanned by
 $\H_i$, and thus the feedback only conveys this information.

\subsection{Finite Rate Feedback Model}
\label{ssec:fbmodel}

 The quantization codebook used at each receiver is fixed beforehand
 and is known to the transmitter and each receiver.
A quantization codebook $\C$ consists of $2^B$ matrices in
$\Complex^{M \times N}$ i.\@e.\@ $(\W_1, $ $\dots, \W_{2^B})$, where
$B$ is the feedback bits per user. The quantization of a channel
matrix $\H_i$, say $\HH_i$, is chosen from the codebook $\C$
according to:
\begin{equation}
\HH_i = \mathop{\arg \min}\limits_{\W\ \in\ \C}\ d^2\LB\H_i, \W\RB
\end{equation}
where $d\LB\H_i, \W\RB$ is the distance metric. Here, we consider the \textit{chordal distance} \cite{CONW96}:
\begin{equation}
d\LB\H_i, \W\RB = \sqrt{\sum\limits_{i=1}^N \sin^2\theta_i}
\end{equation}
where the $\theta_i$'s are the principal angles between the two subspaces spanned by the columns of the
 matrices. As the principal angles depend only on the subspaces spanned by the columns of the matrices,
  it can be assumed that the elements of $\C$ unitary matrices. No channel
  magnitude information is fed back to the transmitter.

\subsection{Random Quantization Codebooks}
\label{ssec:randcode}

Since the design of optimal quantization codebooks for the given
distance metric is a very difficult problem, we instead study
performance averaged over \textit{random} quantization codebooks.
The Grassmannian manifold is the set of all $N$ dimensional
subspaces in an $M$ dimensional Euclidean space, and is denoted by
$\Grassmann$.  Each of the $2^B$ matrices making up the random
quantization codebook is chosen independently and uniformly
distributed over $\Grassmann$, and each matrix can be assumed to be
unitary (points in $\Grassmann$ are equivalence classes of
orthonormal matrices in $\Complex^{M \times N}$). We analyze the
performance
 averaged over all possible random codebooks. The distortion or error associated with a given codebook
 $\C$ for the quantization of $\H \in \Complex^{M \times N}$ is defined as:
\begin{equation}
D = E \LSB d^2(\H,\HH) \RSB = E \LSB \mathop{\min\limits_{\W \in \C}}\ d^2(\H,\W) \RSB
\end{equation}
where $\HH$ is the quantization of $\H$. It is shown in \cite{DAI06} $D$ satisfies: 
\begin{equation} \label{eqn:D}
D\ \leq\ \frac{\Gamma(\frac{1}{T})}{T} (C_{MN})^{-\frac{1}{T}} 2^{-\frac{B}{T}} + N \exp\LSB-(2^BC_{MN})^{1-a}\RSB = \overline{D}
\end{equation}
for a codebook of size $2^B$, where $T = N (M - N)$ and $a \in (0, 1)$ is a real number between $0$ and $1$ chosen such that $\LB C_{MN}2^B \RB^{-\frac{a}{T}} \leq 1$. $C_{MN}$ is given by $\frac{1}{T!}\ \prod\limits_{i = 1}^N\ \frac{(M - i)!}{(N - i)!}$. The second (exponential) term in (\ref{eqn:D}) for the expression of $D$ can be neglected for large $B$.

\subsection{Block Diagonalization}
\label{ssec:bdiag}

The Block Diagonalization strategy when perfect CSI is available at the transmitter involves precoding the signals to be transmitted in order to suppress interference at each user due to all other users (but not due to different antennas for the same user). If $\s_i \in \Complex^{N \times 1}$ contains the $N$ complex symbols intended for the $i^\text{th}$\ ($1 \leq i \leq K$) user and $\V_i \in \Complex^{M \times N}$ is the precoding matrix, then the transmitted vector is given by:
\begin{equation}
\x = \sqrt{\frac{P}{K}} \sum\limits_{i = 1}^K\ \V_i\s_i
\end{equation}
and the received signal at the $i^\text{th}$ user is given by:
\begin{equation} \label{eqn:rxbd}
\y_i = \sqrt{\frac{P}{K}} \H_i^H \V_i\s_i + \sqrt{\frac{P}{K}} \sum\limits_{j = 1, j \neq i}^K\ \H_i^H\V_j\s_j + {\bf n}_i
\end{equation}

It is assumed that a uniform power allocation strategy among users is employed (due to absence of channel magnitude information at the transmitter). Furthermore, in order to maintain the power constraint it is assumed that $\V_i^H\V_i = {\bf I_N}$ and $E[||\s_i||^2] \leq 1$.

Following the BD strategy, each $\V_i$ is chosen such that $\H_j^H\V_i$ is ${\bf 0},\ \forall i \neq j$. This amounts to determining an orthonormal basis for the null space of the matrix formed by stacking all $\H_j, j \neq i$ matrices together. This reduces the interference terms in equation (\ref{eqn:rxbd}) to zero at each user. This is different from Zero Forcing where each complex symbol to be transmitted to the $m^\text{th}$ antenna (among the $N$ antennas) of the $i^\text{th}$ user is precoded by a vector that is orthogonal to all the columns of $\H_{j \neq i}$ as well as orthogonal to all but the $m^\text{th}$ column of $\H_i$.

However, perfect knowledge of the $\H_i$'s at the transmitter is required for zero interference. When finite rate feedback is employed, each $\V_i$ is chosen such that $\HH_j^H\V_i = {\bf 0}\ \forall i \neq j$ which is $\neq \H_j^H\V_i$ in general, and leads to a loss in throughput.

\section{Throughput Analysis}
\label{sec:tanal}

\subsection{Fixed Feedback Quality}
\label{ssec:ffqual}

In the case of perfect CSIT and BD, the transmitter has the ability
to suppress all interference terms giving a per user ergodic
capacity of:
\begin{equation} \label{eqn:R_BD}
R_{BD}(P) = E_\H \LSB\ \log_2 \left| {\bf I_N} + \frac{P}{K}\ \H^H\V_{BD}\V_{BD}^H\H \right|\ \RSB
\end{equation}
where $\V_{BD}$ is the precoding matrix chosen by the BD procedure given the channels of all the users.
The expectation is carried out over all channels $\H$.

For finite rate feedback of $B$ bits per user, multiuser
interference cannot be perfectly canceled and leads to additional
noise power.  Taking this interference into account, the per user
throughput is:
\begin{eqnarray} \label{eqn:R_FB}
R_{FB}(P) = E_{\H_i, \C} \LSB \log_2 \left| {\bf I_N} + \frac{P}{K} \sum\limits_{j = 1}^K\ \H_i^H\V_j\V_j^H\H_i \right| \RSB - \nonumber \\
E_{\H_i, \C} \LSB \log_2 \left| {\bf I_N} + \frac{P}{K} \sum\limits_{j = 1, j \neq i}^K \H_i^H\V_j\V_j^H\H_i \right| \RSB
\end{eqnarray}
where the expectation is carried out over all channels as well as random codebooks ($i$ is any user between $1$ and $K$).

\begin{thm} \label{thm:1}
The rate loss per user incurred due to finite rate feedback with respect to perfect CSIT using Block Diagonalization can be bounded from above by:
\begin{eqnarray*}
\Delta R(P) & = & \LSB R_{BD}(P) - R_{FB}(P) \RSB \\
& \leq & N\ \log_2 (1 + P D)
\end{eqnarray*}
\end{thm}

\begin{proof}
$\Delta R(P) = \LSB R_{BD}(P) - R_{FB}(P) \RSB$
\begin{eqnarray*}
& \mathop{\leq}\limits^{\text{\tiny{(a)}}} & E_\H \LSB \log_2 \left| {\bf I_N} + \frac{P}{K}\ \H^H\V_{BD}\V_{BD}^H\H \right|\ \RSB - \\
& & E_{\H_i, \C} \LSB\ \log_2 \left| {\bf I_N} + \frac{P}{K}\ \H_i^H\V_i\V_i^H\H_i \right|\ \RSB + \\
& & E_{\H_i, \C} \LSB\ \log_2 \left| {\bf I_N} + \frac{P}{K} \sum\limits_{j = 1, j \neq i}^K\ \H_i^H\V_j\V_j^H\H_i \right|\ \RSB \\
& \mathop{=}\limits^{\text{\tiny{(b)}}} & E_{\H_i, \C} \LSB\ \log_2 \left| {\bf I_N} + \frac{P}{K} \sum\limits_{j = 1, j \neq i}^K\ \H_i^H\V_j\V_j^H\H_i \right|\ \RSB\\
& \mathop{=}\limits^{\text{\tiny{(c)}}} & E_{\H_i, \C} \LSB \log_2 \left| {\bf I_N} + \frac{P}{K} \tilde{\H}_i^H \LB \sum\limits_{j \neq i} \V_j\V_j^H \RB \tilde{\H}_i {\bf \Lambda}_i \right| \RSB\\
& \mathop{\leq}\limits^{\text{\tiny{(d)}}} & \log_2\ \left| {\bf I_N} + \frac{P(K-1)}{K} E_{\H, \C} \LSB \tilde{\H}_i^H\LB\V_j\V_j^H\RB\tilde{\H}_i \RSB M\right|\\
\end{eqnarray*}

Here, bound (a) follows by neglecting the positive semi definite interference terms. Both $\V_{BD}$ and $\V_i$ are uniformly distributed and independent of $\H_i$ which results in (b). We write $\H_i\H_i^H = \tilde{\H}_i {\bf \Lambda_i} \tilde{\H}_i^H$ where $\tilde{\H}_i$ forms an orthonormal basis for the subspace spanned be the columns of $\H_i$ and ${\bf \Lambda_i} = \text{diag}[\lambda_1, \dots, \lambda_N]$ are the $N$ non-zero unordered eigen values of $\H_i\H_i^H$ (assuming $\H_i$ is of rank $N$ and diagonalizable) where $E\LSB{\bf \Lambda_i}\RSB$ is $M{\bf I}_N$, and (c) follows. The bound (d) follows from Jensen's inequality due to the concavity of $\log|\cdot|$. It can also be shown that $E_{\H, \C} \LSB \tilde{\H}_i^H\LB\V_j\V_j^H\RB\tilde{\H}_i \RSB = \frac{D}{M - N}$, which provides a bound on the rate loss per user. $D$ can be upper bounded by $\overline{D}$ from (\ref{eqn:D}) for large enough $B$.
\end{proof}

\subsection{Increasing Feedback Quality}
\label{ssec:incfqual}

\begin{thm} \label{thm:2}
In order to maintain a rate loss $\Delta R(P)$ of no larger than $\log_2(b) > 0$ per user, it is sufficient for the number of feedback bits per user to be scaled with SNR as:
\begin{eqnarray} \label{eqn:B}
B \approx & \frac{N (M - N)}{3}P_{dB} - N (M - N) \log_2(b^{\frac{1}{N}}-1)\ + \nonumber\\
& N (M - N) \log_2 \LSB \frac{\Gamma(\frac{1}{N(M-N)})}{N(M-N)} \RSB - \log_2(C_{MN})
\end{eqnarray}
\end{thm}

This expression can be found by equating the upper bound on rate
loss with $\log_2 b$ and solving for $B$ as a function of $P$. Solving this numerically will yield the number of bits strictly sufficient for a maximum rate loss of $log_2 b$. We assume that $B$ is large enough to neglect the exponential term in the expression for $\overline{D}$ from (\ref{eqn:D}) which yields the above approximation. The total contribution of the term containing the logarithm of the gamma function is less than a bit and it can usually be neglected. To maintain a system throughput loss of $M$ bps/Hz, which corresponds to an SNR gap of no
more than $3$ dB with respect to BD with perfect CSIT, it is sufficient to scale the bits as:
\begin{equation} \label{eqn:BDSca}
B \approx \frac{N (M - N)}{3}P_{dB} - \log_2(C_{MN})
\end{equation}

The factor of $N (M-N)$ suggests that the number of feedback bits per antenna reduce with increasing $N$. The number of bits can grow very large for MIMO broadcast systems, and simulation becomes a computationally complex task. However, utilizing the statistics of random codebooks, systems with a small number of antennas can be simulated in a reasonable amount of time. We present simulation results for $M=8$ and $N=2$ in Figure \ref{fig:res1}(a) while scaling the bits as per (\ref{eqn:BDSca}). As Theorem \ref{thm:2} only provides the sufficient number of bits, this is a conservative strategy and the actual SNR gap is found to be $2.3$dB instead of $3$dB. The simulations also suggest that keeping the number of bits fixed will result in rate loss which increases with SNR. Similar results are presented in Figure \ref{fig:res1}(b) for an $N = 3$ system.

\section{Zero forcing vs.\@ Block diagonalization}
\label{sec:ZF}

Zero forcing is an even simpler strategy than BD, and it is
important to compare the performance of these two schemes under the
presence of limited feedback.  Zero forcing for a MIMO broadcast
system with $K$ users and $N$ antennas per user is equivalent to a
$KN = M$ user system with a single antenna per user. The feedback
scaling law for such a system is derived in \cite{JIND05} to be:
\begin{equation} \label{eqn:ZFSca}
B_{ZF} \approx \frac{(M - 1)}{3}P_{dB}
\end{equation}
to maintain an SNR gap of no more than $3$ dB with respect to ZF under perfect CSIT conditions. In general, BD achieves a higher sum rate than ZF with perfect CSIT where the rate gap is $Klog_2(e)$ $\sum_{j=1}^N\frac{N-j}{j}$ \cite{LEE06} at high SNR. In order to compare the number of bits required for BD and ZF under imperfect CSIT and finite rate feedback, it is necessary to fix a common target rate. The bits required per user for ZF must also be multiplied by $N$ for
  fair comparison. By setting $b = 2^{R_g + R}$ in (\ref{eqn:B}) where $R_g$ is the per user rate gap between BD
    and ZF with perfect CSIT and $R$ the target per user rate loss for the ZF system, we can
    compare the \textit{sufficient} number of bits required to achieve the same sum rate for
     both strategies. For example, $R = 1$ for a $3$ dB target and this suggests a  bit
      savings of 20\% for an $M = 6, N = 2$ system, and 25\% for an $M = 9, N = 3$ system with BD. The scaling law in Theorem \ref{thm:2} is however highly conservative for large $b$, and though it is possible to see that BD has a clear
       advantage in terms of the sufficient number of bits required, it is somewhat underestimated.
       If a ZF system is scaled to maintain a $3$dB SNR gap relative to perfect CSIT and the number
        of feedback bits for BD is numerically determined to achieve the same sum rate, bit savings
         are about 40-50\% for an $M = 6, N = 2$ system.

\section{Quantizing the Channel}
\label{sec:scaquan}
The scaling law in Section \ref{ssec:incfqual} was derived considering random codebooks, which are impractical for real world applications. Although vector quantization codebooks can be designed for more practical systems it is likely to require very high complexity due to the large number of bits at each mobile. It is thus worthwhile to investigate low complexity scalar quantization schemes. We believe that simple scalar quantization methods are capable of achieving the same bit scaling rate as random codes, though they will incur a constant rate loss.

The scalar quantization scheme is first presented for MISO systems (based on the idea in \cite{Naru98}). A complex channel vector $\H_i = [H_1, \dots, H_M]^H \in \Complex^{M \times 1}$ is first divided by one of its elements, say $H_1$, to yield $M - 1$ complex elements. The phase of each of these elements is quantized separately and uniformly in the interval $[-\pi, \pi]$. The inverse tangents of the magnitudes, for example $\tan^{-1}\LB\frac{|H_2|}{|H_1|}\RB$, are quantized uniformly in the interval $[0, \frac{\pi}{2}]$. Nonuniform quantization based on the distribution of these random variables is also possible, but (sub-optimal) uniform quantization appears to be sufficient for the number of feedback bits to scale linearly with SNR with the same slope as with random codebooks. The total number of bits available to a user is assumed to be distributed equally among the phases and magnitudes of the $M-1$ elements as far as possible, and the remaining bits are randomly assigned.

For MIMO systems, this is generalized to quantizing the magnitude and phase (in the same manner) of the lower $(M-N) \times N$ entries of the matrix $\H_i([{\bf I}_1\ |\ {\bf Z}_1]\H_i)^{-1}$. ${\bf I}_1$ is the $N \times \frac{M}{2}$ identity matrix and ${\bf Z}_1$ the $N \times \frac{M}{2}$ zero matrix.

\begin{figure}[ht]

\begin{minipage}[ht]{1.0\linewidth}
\centering
\centerline{\epsfig{figure=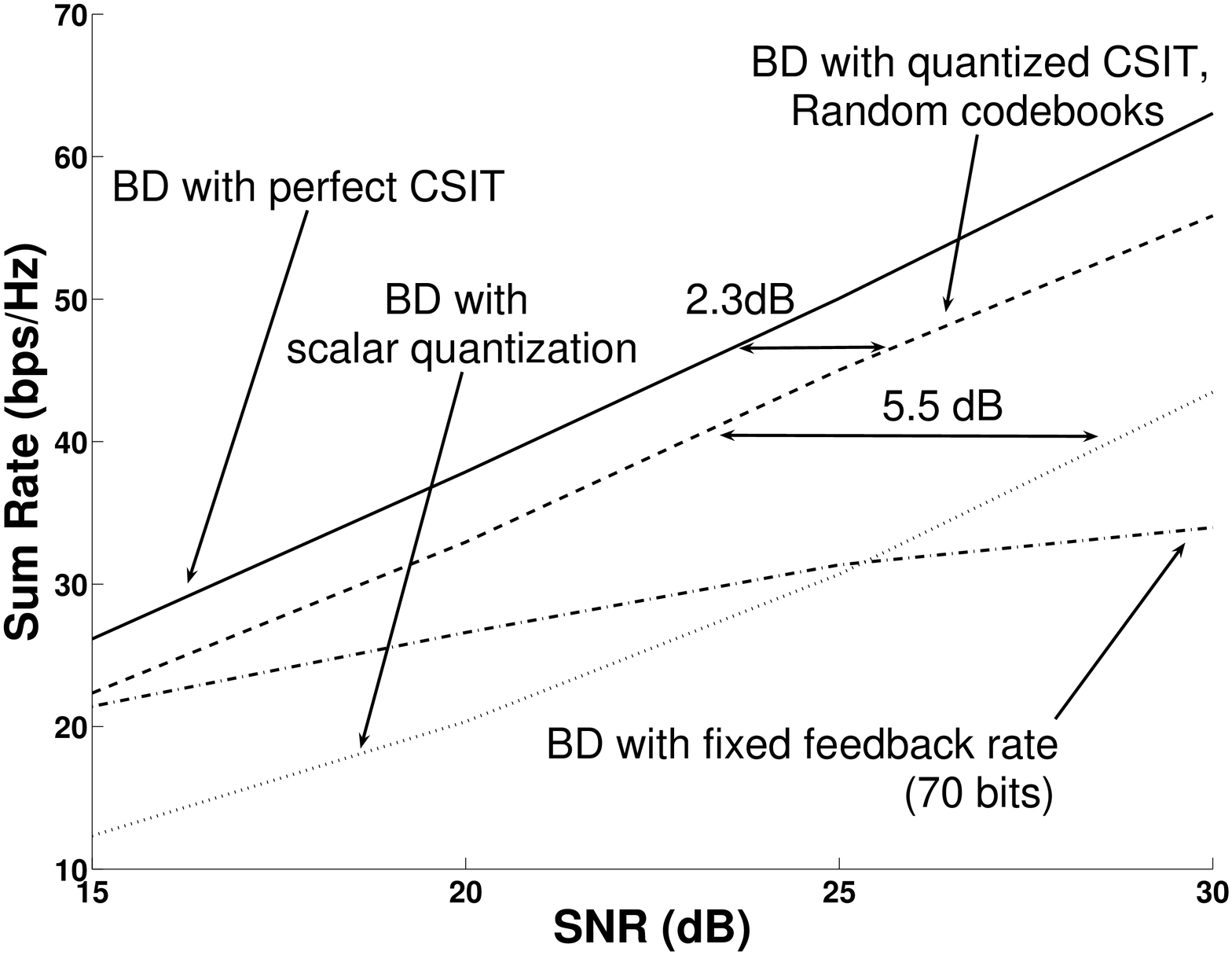,width=8.5cm}}
\centerline{(a) MIMO Broadcast Channel with M = 8, N = 2, K = 4}\medskip
\end{minipage}
\begin{minipage}[ht]{1.0\linewidth}
\centering
\centerline{\epsfig{figure=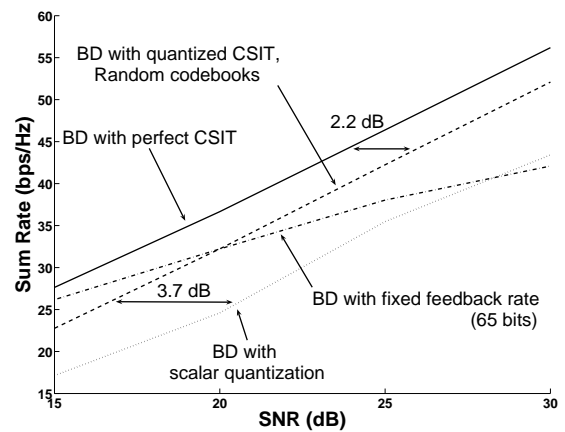,width=8.5cm}}
\centerline{(b) MIMO Broadcast Channel with M = 6, N = 3, K = 2}\medskip
\end{minipage}

\caption{Sum Rate with Block Diagonalization and finite rate feedback}
\label{fig:res1}
\end{figure}

\begin{figure} [hb]
\begin{minipage}[b]{1.0\linewidth}
\centering
\centerline{\epsfig{figure=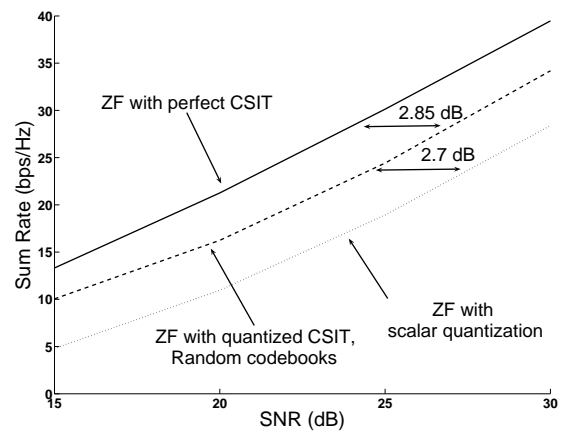,width=8.5cm}}
\end{minipage}
\caption{Scalar quantization in MISO systems (M = 6, N = 1)}
\label{fig:res2}
\end{figure}

Although we do not offer an analytical proof that this scheme achieves the same bit scaling rate as random codebook quantization, we present simulation results that certainly suggest this. The bits for scalar quantization are scaled according to Equation (\ref{eqn:ZFSca}) for an $M=6, N=1$ MISO system (Figure \ref{fig:res2}). This maintains a constant gap with the perfect CSIT curve, although there is a $2.7$ dB SNR loss with respect to random codebook quantization. More sophisticated scalar quantization methods may be able to reduce this gap as well, which indicates that simple scalar quantization schemes could perform quite well. Similar results for MIMO systems are presented in Figures \ref{fig:res1}(a) and \ref{fig:res1}(b) with $4$ and $2$ users respectively.


\bibliographystyle{hieeetr}
\bibliography{strings}

\end{document}